\documentclass[prc,showpacs,floatfix,twocolumn,superscriptaddress,nofootinbib]
{revtex4}
\usepackage{colordvi}
\usepackage{color}
\usepackage{graphics}
\usepackage{longtable}
\usepackage{epsfig}
\usepackage{amsmath,amssymb,amsfonts}

\begin{document}




\title{QRPA treatment of the transverse wobbling mode  reconsidered }

\vskip 10pt
\author{S. Frauendorf}
\email{sfrauend@nd.edu} 
\affiliation{Department of Physics, University of Notre Dame, South Bend, Indiana 46556}
\author{F. D\"onau$\dagger$}
\begin{abstract}
{\bf Abstract}\\
The quasiparticle random phase approximation is used to study the properties of the wobbling bands in $^{163}$Lu. 
Assuming that the wobbling mode represents pure isoscalar orientation oscillations  results in too low wobbling frequencies and
too strong  M1 transitions between the one- and zero-phonon  wobbling bands. 
The inclusion of an LL interaction, which couples the wobbling mode  to the scissors
mode,  generates  the right upshift of the wobbling frequencies and the right 
suppression of the  B(M1)$_{out}$ values  toward the experimental values. 
In analogy to the quenching of low-energy E1 transition by coupling to the Isovector Giant Dipole Resonance,   a general reduction 
of the M1 transitions between quasiparticle configurations caused by coupling to the scissors mode is suggested. 

\end{abstract}


\maketitle

\section{INTRODUCTION}
Rotating nuclei that have a triaxially deformed shape  are expected exhibit a characteristic excitation mode called "wobbling" by Bohr and Mottelson \cite{Bo75}, which is an 
orientation vibration of the triaxial body about  the  rotational axis.  It is the nuclear analog  to the motion of the classical top with three different moments of inertia, which
is well known from the rotational spectra of molecules. Experimental evidence for the 
wobbling mode was established by the discovery \cite{Od01,Je01,Je02} of rotational bands in the $^{71}$Lu isotopes when they attain a 
triaxial strongly deformed (TSD) shape at high spin. The simple dynamics of a rotor with three different moments of inertia results in an increase
of the wobbling frequency with angular momentum, which is seen in molecules. However, for the Lu isotopes a decrease is observed, which makes the identification of the wobbling 
possible, because it prevents the mode being fragmented among competing quasiparticle excitations. In the framework of the Quasiparticle+Triaxial-Rotor (QTR) model, 
Frauendorf and D\"onau \cite{FD14} demonstrated that the decrease results from the presence of the odd i$_{13/2}$ quasi proton, which aligns its angular momentum along the
short body axis, {\it transverse} to the medium axis with the largest moment of inertia.   To notify  the modification of the dynamics by the odd quasiparticle, 
they introduced the  name "transverse wobbling". They predicted the  appearance of transverse wobbling for the mass 130 region, where the h$_{11/2}$ quasiparticle couples
transverse to the triaxial rotor. The prediction was recently confirmed for $^{135}$Pr \cite{135Pr}.  The QTR calculations well account for the wobbling energies and the 
B(E2)$_{con}$ values of the $\Delta I=1$ electric quadrupole transitions, which connect the one-phonon wobbling band with the zero-phonon band. 
However, the B(M1)$_{con}$ values of the connecting magnetic  dipole transitions are overestimated by about a factor of 3-10 (see Ref. \cite{135Pr} and Ref. \cite{FD14}).   
The discrepancy turns out to be robust, and it can be traced back to the transverse geometry: For a quasiproton that is rigidly coupled to the triaxial charge density distribution
(HFA approximation of Ref. \cite{FD14}) the amplitude of the wobbling vibrations of the charge density, which generate the B(E2)$_{out}$ values of the inter band transitions, determines the 
the amplitude of the vibrations of the magnetic moment of the odd quasi proton, which generate the B(M1) values of the inter band transitions. Realistically, the odd quasi proton
is not rigidly coupled to the rotor, which reduces the amplitude of the oscillations of the magnetic moment and thus the B(M1)$_{out}$ values. However the reduction is too weak to bring 
down the B(M1)$_{out}$  to the experimental values (see Fig.  19 of Ref. \cite{FD14}). 
The present paper addresses this
problem of the too strong magnetic dipole transitions from a microscopic perspective. 

 Following the  discovery of the first wobbling structure in $^{163}$Lu~\cite{Od01}, Hamamoto, Hagemann  {\it et al.} \cite{Od01,Ha02,Ha03} used the QTR model to describe the 
wobbling mode. These calculations made the ad hoc assumption that the short axis has the largest 
moment of inertia, by exchanging the hydrodynamic moments of inertia of the short and medium axes.
The large ratios B(E2)$_{out}$/B(E2)$_{in}$  of inter-band to intra-band  E2 transitions  could be well reproduced. The B(M1)$_{out}$
were only overestimated by a factor of 2-3. However,  the calculated wobbling frequencies of the  QTR model with the "inverted moments of inertia" assumption
distinctly disagree with experiment. Instead of the experimentally observed decrease, the 
wobbling frequency increases with the spin $I$, which is expected because the inverted moment of inertia arrangement corresponds to the longitudinal wobbling geometry 
in the terminology of Ref. \cite{FD14}. Ref.  \cite{Ta10} suggested to remedy the problem by assuming a decrease of the scale of the rotational energy, which may reflect the increase
of the moments of inertia due to a reduction of the pair correlations. In our view, the  "inverted moments of inertia" assumption is unrealistic because any microscopic calculation
of the three moments of inertia in the frame of the cranking model give the maximal moment of inertia for the medium axis. This result is in accordance with the
hydrodynamic ratios between the moments of inertia. It can be qualitatively understood by the fact that the moment of inertia of a certain axis increases with the deviation from  
cylindric symmetry, which is maximal for the medium axis. Hence, the problem with the too strong magnetic transition remains.

The observation of the wobbling mode stimulated theoretical efforts to understand how the nuclear shell structure and the residual interaction 
generate such a type of collective excitations. Matsuyanagi, Matsuzaki, Ohtsubo, Shimizu, and Shoji demonstrated that
  the quasiparticle random phase approximation (QRPA) is an 
 adequate microscopic approach \,\cite{Sh95a,Ma02,Ma04b,Shoji,Sh08}. QRPA 
  describes wobbling bands in terms of correlated two-quasiparticle excitations in a rotating triaxial potential. Relevant results of these 
studies can be summarized as follows. \\
- The QRPA calculations agree with the transverse wobbling geometry as discussed in Ref. \cite{FD14}. The authors refers to it as 
"positive $\gamma$ shape", which uses the common terminology of principle axis cranking that assigns the sector $0\leq\gamma \leq 60^\circ$ to rotation 
about the short axis. The angular momentum of the odd i$_{13/2}$ quasiparticle aligns with this axis. The decrease of the wobbling frequency 
is interpreted as the approach of the instability of the cranking solution to a tilt of the rotational axis into the short-medium plane, which is signaled by 
the frequency of the lowest QRPA solution to become zero \cite{Ma04b}.  \\
-The collective enhancement of the connecting E2 transitions is born out. QRPA calculations based on the Niisson potential underestimate the
ratios B(E2)$_{out}$/B(E2)$_{in}$  by about a factor of two \cite{Sh95a,Ma02,Ma04b}, the ones based on a Woods-Saxon potential get it right \cite{Shoji}. \\
-The B(M1)$_{out}$ values of the inter band transitions are overestimated by a factor of 10 as for the QTR results for transverse wobbling. 

The QRPA calculations \cite{Sh95a,Ma02,Ma04b} used  an isoscalar quadrupole-quadrupole (QQ) residual interaction.
 Because such interaction generates the same coupling between the odd quasiparticle and the triaxial rotor core as  
  in the QTR calculations, it comes as no surprise that both approaches overestimate    B(M1)$_{out}$ values by the same factor.
The  reason to revisit the QRPA in this paper is to investigate how modifying the residual interaction influences the resulting excitation energies and electromagnetic transition
  rates. In particular we are interested whether  the suppression of the inter band  M1 transitions can be obtained for transverse wobbling.
  We study  the i$_{13/2}$ TSD bands in $^{163}$Lu which offer the most complete set of data. 

The paper is organized as follows.
In Sec.\ref{sec:ISQQsc} a selfconsistent treatment of the QRPA is performed by deriving the shape parameters $(\varepsilon, \gamma)$  from the QQ interaction.
In Sec.\ref{sec:ISQQsr} the shape parameters $(\varepsilon,\gamma)$ are adopted from a Nilsson-Strutinsky minimization \cite{Go04} 
and the strength of residual QQ interaction is determined by restoring the rotational 
invariance of the Hamiltonian. Sec.\ref{sec:AddRI} studies the consequences of additional interactions.
 Coupling to the low-energy orbital M1 resonance ("scissors mode") is suggested as a mechanism that suppresses the strength of the M1 inter band transitions.
Sec. \ref{sec:SC}  summarizes the results and puts them into perspective.

\section{Quasiparticle random phase approximation (QRPA) for isoscalar QQ interaction }\label{sec:ISQQ}
\subsection{Selfconsistent QRPA (sc QRPA) with standard QQ interaction}\label{sec:ISQQsc}
The theoretical framework of our QRPA calculations is  similar to the one used in
our recent study  of chiral  vibrations \cite{Alme11}.
The Hamiltonian $\hat{H'}$ is defined with respect to a  reference system rotating about the 1-axis,
\begin{eqnarray}  
\label{eq:H1} 
\hat{H'}=\hat{H}- {\omega} {\hat{J}_1} ,
\end{eqnarray}
where $\omega$ is the cranking frequency and $\hat{J}_1$ denotes the 1-component  of the angular momentum operator. 
The cranking term $-\,\omega \,\hat{J}_1$ ensures that the states have an average angular momentum $\langle J_1\rangle=I $.
The corresponding lab. Hamiltonian $\hat{H} $ in Eq.(\ref{eq:H1}) is 
\begin{eqnarray}
\label{H2} 
\hat{H}  &=& 
\sum_{\tau =\pi,\nu}
 [\,\,\hat{h}^\circ_{\tau}   
-\Delta_\tau (\hat{P}^\dagger_\tau + \hat{P}_\tau)  
- \lambda_\tau \hat{N}_\tau  \,\,]
 \nonumber\\
&&\quad\quad -\frac{\kappa_{_0}}{2} 
\sum_{m=-2,2} (-1)^{m}\hat{Q}_{m}\hat{Q}_{-m}. 
\end{eqnarray} 
The operator $\hat{h}^\circ_{\tau}$ is the spherical part of the Nilsson Hamiltonian where the
isospin index $\tau=\pi,\nu$ distinguishes the neutron and proton contributions, respectively. 
The term $\Delta_\tau ( \hat{P}^\dagger_\tau + \hat{P}_\tau )$ accounts for the pair field 
where $\hat{P}^\dagger_\tau$ and $\hat{P}_\tau $ are the familiar monopole pair operators.
Aiming at the high-spin $\pi i_{13/2}$ band in $^{163}$Lu,
the gap parameters $\Delta_\tau$  are assumed to be reduced:
below the cranking frequency $\omega$=0.45 MeV  we take $\Delta_{\pi}$=0.45 MeV for the proton
gap and $\Delta_{\nu}$=0.35 MeV for the neutron gap, and we use  $\Delta_{\tau=\pi,\nu}$$=$$0$ above.
As usual, the terms $\lambda_\tau \hat{N}_\tau$,  containing the particle
number operators $\hat{N}_\tau$, are introduced to attain the 
average particle numbers $\langle \hat N_{\pi}\rangle=Z$
and $\langle \hat N_{\nu}\rangle=N$, respectively, by an 
appropriate choice of the Fermi energy $\lambda_\tau$. 
The following term in Eq.\,(\ref{H2}) is the isoscalar quadrupole-quadrupole  (ISQQ) interaction 
constructed from the mass quadrupole operators $\hat Q_{m} = \hat Q_{m}(\pi)$+ $\hat Q_{m}(\nu)$ where $\hat Q_{m}(\tau)\equiv \sqrt{4\pi/5}\,\,(r/b_\circ)^2 Y_{2m}(\tau)$ 
and $b_\circ= 1.01 A^{1/3}$ is the oscillator length. 

In this section we assume selfconsistency between the 
ISQQ interaction and the deformed nuclear shape
defined by the parameters $(\varepsilon,\gamma)$. More precisely, it is the deformed mean field potential $v$ of the ISQQ interaction which, 
for a predefined interaction strength $\kappa_{_0}$, has to obey the condition 
\begin{equation}
\label{Qmf}
         v=v(\varepsilon,\gamma) =-\kappa_{_0} \lbrack \langle \hat Q_{_0}\rangle \hat Q_{_0}+ 
\langle \hat Q_{_2}\rangle(\hat Q_{_2}+\hat Q_{_{-2}})\rbrack,
\end{equation}
where $|\rangle = |\,\varepsilon,\gamma\rangle$ is the quasiparticle reference state of  the $\pi$i$_{13/2}$ TSD band as specified below. 
\begin{table}[t]  
\caption{   
Equilibrium values  of the deformation parameters $(\varepsilon, \gamma)$ in the frequency region $\omega=$
0.15-0.50 MeV/$\hbar$. The strength parameter of the 
ISQQ interaction is $\kappa_{_0}$= 0.01960 MeV} 
\label{tab:defs}
\begin{ruledtabular} 
\begin{tabular}{ccc} 
$\omega$(MeV/$\hbar$) & $\varepsilon$& $\gamma$(deg) \\\hline
      0.15&        0.398839&     9.248 \\
        0.20&       0.397926&     9.362\\
        0.25&      0.396632 &   9.486  \\
        0.30&       0.394788&    9.631\\
        0.35&       0.392064&    9.798\\
        0.40&       0.387658&     9.977\\
        0.45&       0.381236&     11.575\\
        0.50&       0.377065&     11.619\\
\end{tabular} 
\end{ruledtabular} 
\end{table} 
Denoting the c-numbers 
$\langle \hat Q_{_{0,2}}\rangle$  as $q_{_{0,2}}(\varepsilon,\gamma)$ 
the selfconsistency conditions demand searching for deformation parameters which at a given cranking frequency $\omega$ satisfy the relations 
\begin{eqnarray}\label{scepsga}
\kappa_{_0}\langle \hat Q_{_0}\rangle\equiv\kappa_{_0} q_{_0}(\varepsilon,\gamma) 
&=&\quad 2/3\,\hbar\omega_{_0} \,\varepsilon \cos{\gamma},\nonumber\\
\kappa_{_0}\langle \hat Q_{_2}\rangle\equiv\kappa_{_0} q_{_2}(\varepsilon,\gamma) 
&=&-2/3\,\hbar\omega_{_0}\,\varepsilon \sin{\gamma}/\sqrt{2}.\quad\quad
\end{eqnarray}
The mean field calculations are
done by using the tilted axis cranking (TAC) code described in Ref.  \cite{Fr00}.
 It should be noted that the above conditions lead to a stable 
equilibrium shape only if one renders the volume conservation by taking  the scale factor $\hbar\omega_{_0}=41A^{-1/3}$\,MeV as constant. Combining  the spherical mean field part from the 
Hamiltonian  $\hat H'$, Eq.(\ref{eq:H1}), with the 
selfconsistency conditions (\ref{scepsga}), one obtains the mean field Hamiltonian of the 1D-TAC model $\hat{h'}=\,\,\hat{h}  - {\omega} {\hat{J}_1}$ \cite{Fr00}, where $\hat{h}$ is given by
\begin{eqnarray}\label{htac}
\hat{h}=\,\,\hat{h}^\circ  -
\Delta_\tau (\hat{P}^\dagger+ \hat{P} ) - \lambda \hat{N} -\qquad\qquad\qquad\qquad\nonumber\\
- \,\hbar\omega_{_0} \frac{2}{3}\,\varepsilon \,\left(\,\cos{\gamma} \hat Q_{_0} 
 -  \frac{\sin{\gamma}}{\sqrt{2}}(\hat Q_{_2}+\hat Q_{_{-2}})\,\right)
.\quad
\end{eqnarray}
\begin{figure}[htbp] 
\vspace*{-3cm}
\includegraphics[width=13cm]{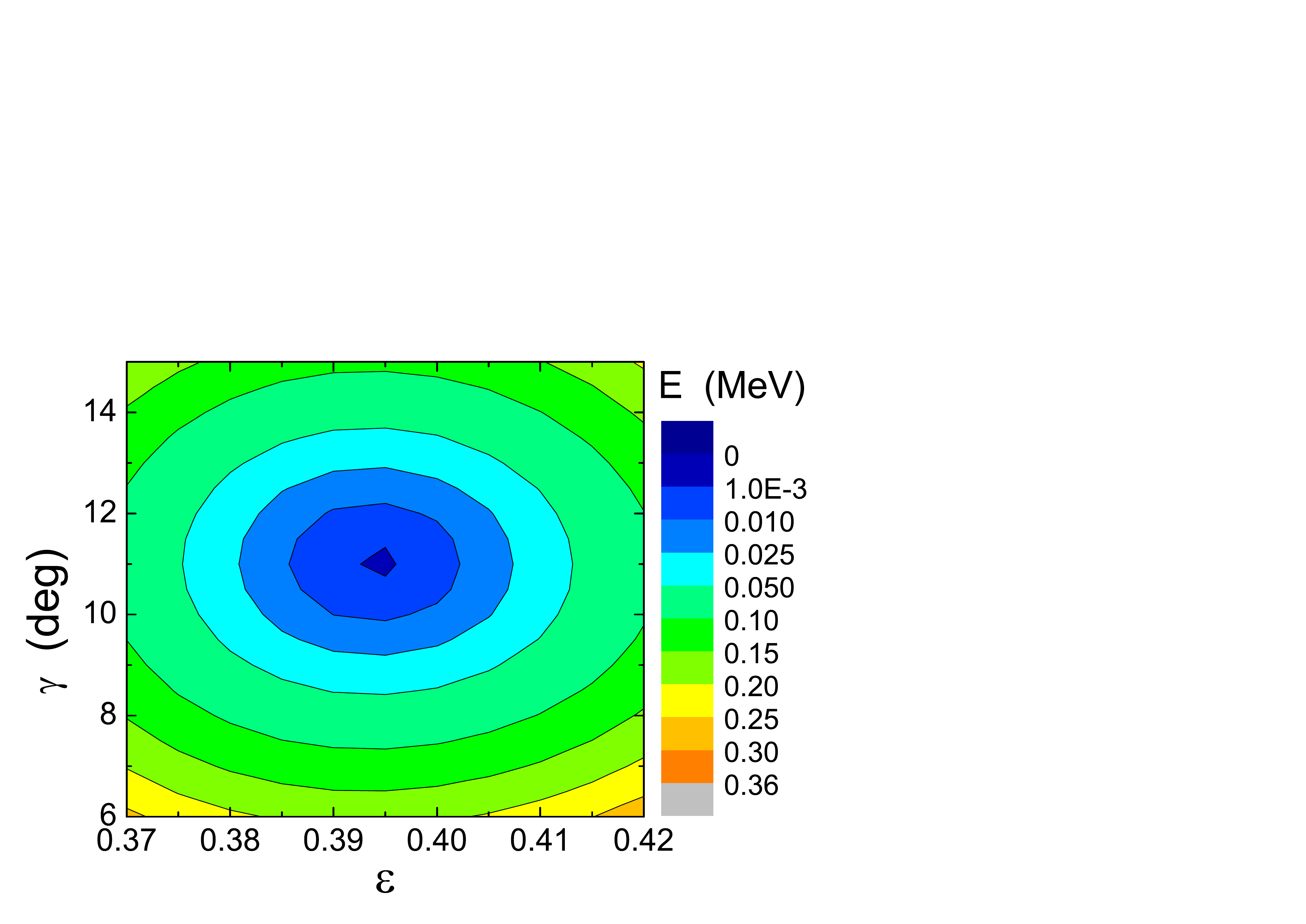}
  \caption{(Color online) Total routhian surface for the TSD configuration in $^{163}$Lu at  $\omega=0.45$ MeV/$\hbar$.} 
  \label{fig:shape} 
\end{figure} 
The diagonalization of the  TAC Hamiltonian $\hat{h}$ is done in an oscillator 
basis with the quantum numbers $\{n, l, j, m\}$ including the orbits of the three main shells $n=4-6$. 
The search for the equilibrium needs to be performed with diabatic tracing (c.f. \cite{Fr00}) of the selected $(\pi $i$_{13/2}, \nu$ g) 
configuration of the TSD band.  
The strength of the sc ISQQ interaction $\kappa_{_0}$=  0.01960 MeV is $\omega$-independent and chosen such that at $\omega$~=~0.15 MeV/$\hbar$  
the deformation parameter comes close to the suggested value $\varepsilon=0.4$
of the experimental TSD band  \cite{Go04}.
\begin{figure}[htbp] 
\includegraphics[clip,width=\linewidth]{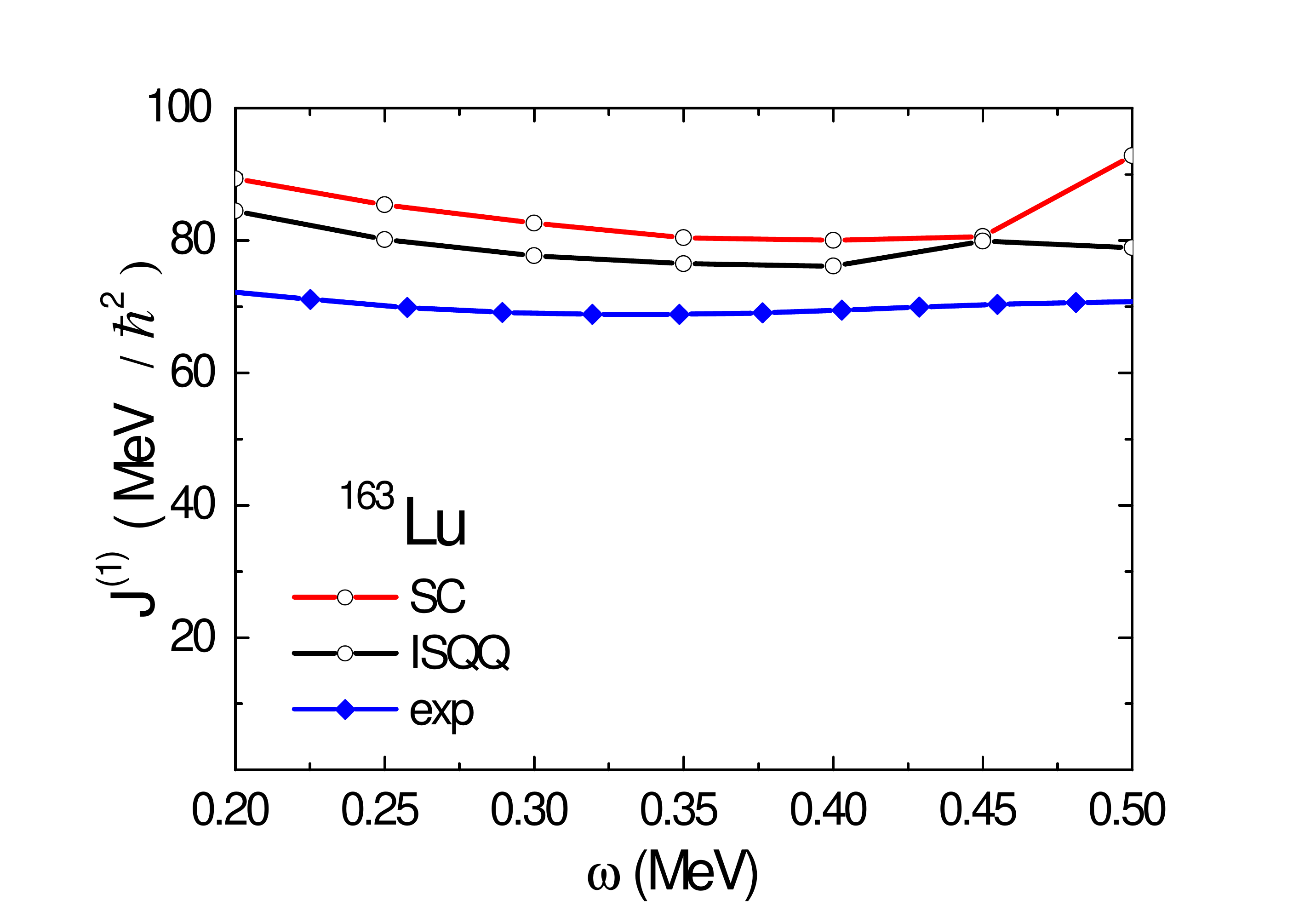} 
 \caption{(Color online) Experimental and calculated kinematic moments of inertia of the
TSD band in $^{163}$Lu. The calculated moment of inertia is ${\cal J}^{(1)}=\langle J_1\rangle /\omega$.} 
  \label{fig:MoI} 
\end{figure}  
The sc deformation parameters  for the frequency interval  $\omega$=0.15-0.50 MeV/$\hbar$ are presented in 
Table\,\ref{tab:defs}. It is seen that for the ISQQ interaction the sc triaxiality parameter $\gamma$ $\approx 9-12 ^\circ$ 
is lower than +20$^\circ$ suggested  in Ref.\,\cite{Go04},  but close to the values found by Shoji and Shimizu \cite{Shoji} with 
Nilsson-Strutinsky  minimization.  
The relative change of the deformation $(\varepsilon, \gamma)$ 
to higher rotational frequencies is small. Nevertheless  precise  selfconsistency is required 
in the subsequent QRPA calculation in order to obtain reliable values for the excitation energies and 
E2/M1 properties of the wobbling band.  As already noted in the previous QRPA papers \cite{Sh95a,Ma02,Ma04b,Shoji,Sh08},  
the absolute minimum of $\langle \hat{H'} \rangle$ 
corresponds to rotation about the short axis of the triaxial potential, along which the angular momentum of 
the i$_{13/2}$ proton is aligned (the sector of positive $\gamma$-values in standard Principle Axis Cranking  (PAC) terminology).
Above the frequency $\omega$=0.5 MeV/$\hbar$ the PAC solution becomes unstable, 
because the moment of inertia of the medium axis is larger than the one of the short axis. The stable solution
corresponds to rotation about a $tilted$ axis in the short-medium plane, which represents a $\Delta I=1$ band.
The QRPA frequency goes to zero when approaching the instability from below.
Thus, the QRPA solution studied in this paper is of  the "transverse wobbling" 
type according to the classification scheme introduced by us in Ref. \cite{FD14}, where the corresponding 
physics is discussed in the semiclassical frame work of the HFA approximation.

\begin{figure}[t] 
\includegraphics[clip,width=\linewidth] {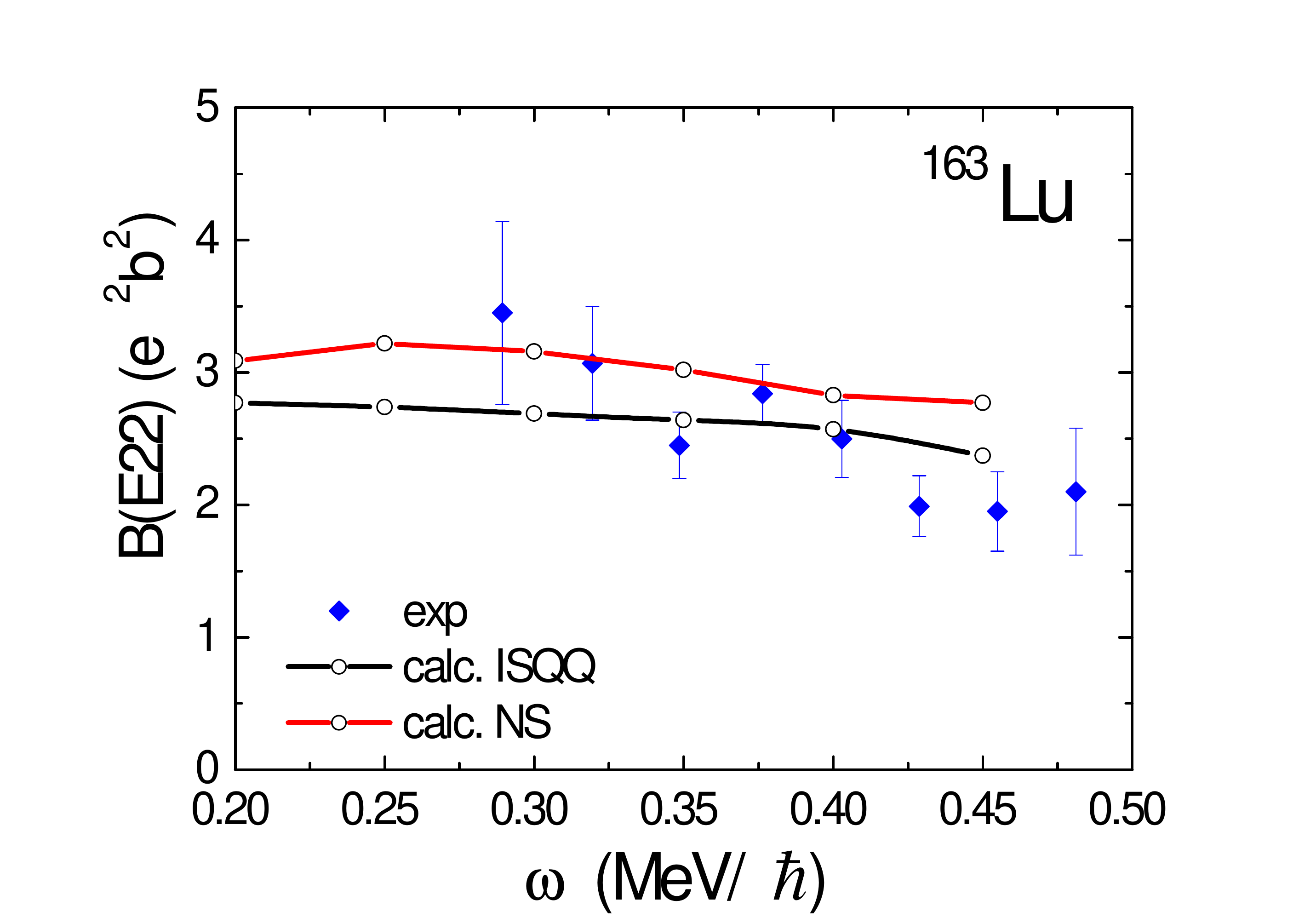}
 \caption{(Color online) Experimental and calculated B(E2,I$\rightarrow $I-2) values of the TSD band.} 
 \label{fig:BE22} 
\end{figure}  
In Fig.\,\ref{fig:shape} we show  the total routhian surface for 
$\omega$=0.45 MeV/$\hbar$ as obtained by diabatic tracing the TSD configuration with the TAC code.
The ISQQ interaction gives a relatively shallow minimum on the deformation surface.
In Fig.\,\ref{fig:MoI} the experimental and calculated moments of inertia ${\cal J}^{(1)}$ are compared. 
The experimental frequencies of the TSD bands are derived by using the standard definition 
$\omega(\bar I)=(E(I)-E(I-2))/2$, where transition spin $\bar I$\,=\,$I-1/2$, and the experimental 
moment of inertia ${\cal J}^{(1)}(\bar I)=\bar I/ /\omega(\bar I)$. The calculation somewhat overestimates the experimental 
values.

Fig.\,\ref{fig:BE22} presents the experimental B(E2) values of the $I\rightarrow\,I-2$
transitions within the TSD g-band \cite{Go04} and the ones calculated 
with the sc TAC model. The polarization charges $e_p$=(1+Z/A)$e$ and $e_n$=Z/A\,$e$ were 
adopted for  the proton and neutron parts of the electric quadrupole operator.

Starting from results of the sc TAC calculation the QRPA is performed following the general formalism 
as outlined in the textbooks (e.g. \cite{RS80}).  We mention only the important 
steps of the QRPA and refer for more details  to our recent paper \cite{Alme11}.
Firstly, the Hamiltonian (\ref{eq:H1}) is rewritten in quasiparticle (qp) representation, 
 \begin{eqnarray} 
 \label{eq:hrpa}
\hat{H'} = \hat{h'}+ \hat{V}_{4qp}, 
 \end{eqnarray}
  where $\hat{h'} $ is the diagonalized TAC Hamiltonian  
\begin{equation} 
  \hat{h'}= E_\circ+ \sum_i e_i \hat{\alpha}_i^\dagger \hat{\alpha}_i . 
\end{equation} 
The set $\{\hat{\alpha}_i^\dagger ,\hat{\alpha}_i\}$ denotes the qp operators,
 $e_i$ are the qp energies  
 and  $\hat{V}_{4qp}$ contains the residual 4qp interaction terms which give rise to the vibrational excitations.
 Then,  the quasi-boson approximation 
$ \hat{\alpha}_i^\dagger \hat{\alpha}_j^\dagger   \Rightarrow \hat{b}_{ij}^\dagger $ is applied 
such that  the Hamiltonian,  Eq.\,(\ref{eq:hrpa}), is expressed in terms of bosons, 
$\hat{H'} 
\Rightarrow 
\hat{H'}_{RPA} $, 
keeping  only boson terms up to second order ~\cite{RS80}. 
This Hamiltonian is diagonalized by  using the QRPA equation 
\begin{equation} 
  \label{eq:rpa1} 
  \left[\hat{H'}_{\rm QRPA}, \;\,\hat{O}^\dagger_\lambda \right]= 
E_{_{QRPA}}^\lambda  \hat{O}^\dagger_\lambda, 
\end{equation}  
which yields the phonon excitation energies $E_{_{QRPA}}^\lambda$ 
and the phonon excitation operators $\hat{O}^\dagger_\lambda$ defined by 
\begin{equation} 
  \label{eq:rpa2} 
  \hat{O}^\dagger_\lambda = \sum_{\mu=i<j} 
(X^\lambda_\mu \hat{b}_\mu^\dagger - 
  Y^\lambda_\mu \hat{b}_\mu).
\end{equation}  
The  amplitudes $X^\lambda_\mu$ and $Y^\lambda_\mu$ are found by solving the standard set of linear 
equations following from Eq.(\ref{eq:rpa1}). The quasiparticle Hamiltonian $\hat h'$ and the full Hamiltonian
$\hat H'$ commute with the signature operator $R_1 =$ exp$(-i\pi \hat I_1)$, which 
generates a 180 deg rotation about the cranking axis. Therefore, the quasiparticle states and the phonon excitations have good signature quantum numbers. 
The energetically lowest phonon state with negative signature $r=-1$ embodies the wobbling excitation which is characterized also by
giving the largest cross-over transition strength B(E2,$I\rightarrow I-1$).
Accordingly, only  2qp components with the combined qp signature $r=r_ir_j=-1$  contribute to the wobbling operator $\hat{O}^\dagger$ in  Eq.(\ref{eq:rpa2}).
One has to make sure that the spurious rotational solution  
with the energy  $E_{_{QRPA}}=\hbar\omega$   does not mix  with the wobbling solution. 
Selfconsistency of the mean field ensures  this requirement.

The E2/M1  transition amplitudes
 from the TSD wobbling band to the TSD g-band  are obtained by evaluating the matrix element
\begin{equation} 
  \label{trans} 
  \langle w|\hat{\cal M}_m(E2/M1)|0\rangle = 
 \langle 0|\hat{O}_w  \hat{\cal M}_m(E2/M1) |0\rangle,
 \end{equation}  
 where $|w\rangle$  means the wobbling phonon state and  $|0\rangle$ denotes the QRPA vacuum state
at the cranking frequency $\omega$. The transition operators are 
\begin{eqnarray}\label{eq:BE2} 
  \hat{\cal M}_m(E2) &=& e_p r_p^2 {Y}_{2m} (p)+e_n r_n^2 {Y}_{m} (n),\\
  \hat{\cal M}_m(M1)& =&\frac{3}{4\pi}
g^{(l)}_p\hat{l}_{1m}(p)+g^{(s)}_p \hat{s}_{1m}(p)+g^{(s)}_n\hat{s}_{1m}(n) . \nonumber
\label{eq:BM1}
\end{eqnarray} 
The component $m$ is assigned to the transition $ I\rightarrow I-m$.
Further, the orbital g-factor for M1 is $g^{(l)}_p= 1 ~\mu_N$ for protons  and 0 for 
neutrons. The spin g-factors $g^{(s)}_p$ and $g^{(s)}_n$ are 0.7 times the values for the 
free proton or neutron. The reduced transition probabilities are   
 \begin{eqnarray}\label{eq:BE2b} 
  B(E2/M1,I\rightarrow I\mp 1) &=& \left|\left< w\left|\hat{\cal M}_{\pm 1}(E2/M1) \right|0 \right>\right|^2.
\quad\quad
 \end{eqnarray} 
\begin{figure}[htbp] 
\includegraphics[clip,width=\linewidth]{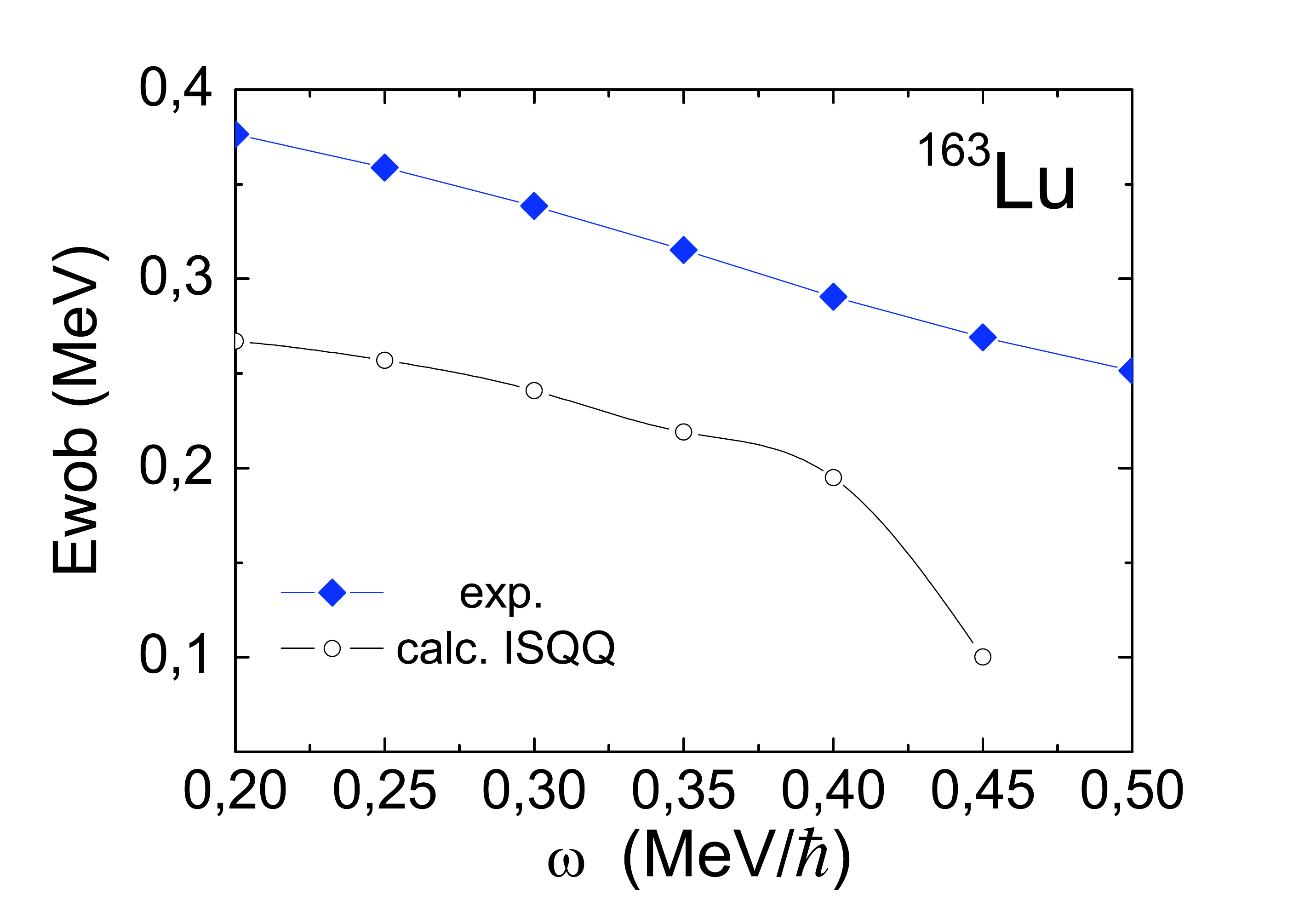}
  \caption{(Color online)  Excitation energy of the wobbling band in $^{163}$Lu as a function of the 
rotational frequency. Experimental values (blue diamonds) are from \cite{Go04}. 
QRPA calculation (solid line) with selfconsistent ISQQ interaction.} 
  \label{ewoba} 
\end{figure} 
In the sc version of QRPA, the ISQQ term in the Hamiltonian (\ref{eq:H1})  generates 
both the deformed mean field and the residual interaction.  As discussed above,  its
 strength  is fixed to the value  $\kappa_\circ=0.01960$ for
the whole frequency range $\omega$ =0.15-0.5 MeV/$\hbar$.
The factorized form of the ISQQ term reduces  the solution of the QRPA equation  to searching the 
zeros  of the dispersion determinant, which are located at  the QRPA energies 
$E_{_{QRPA}}$. 

\begin{figure}[htbp] 
\includegraphics[clip,width=\linewidth]{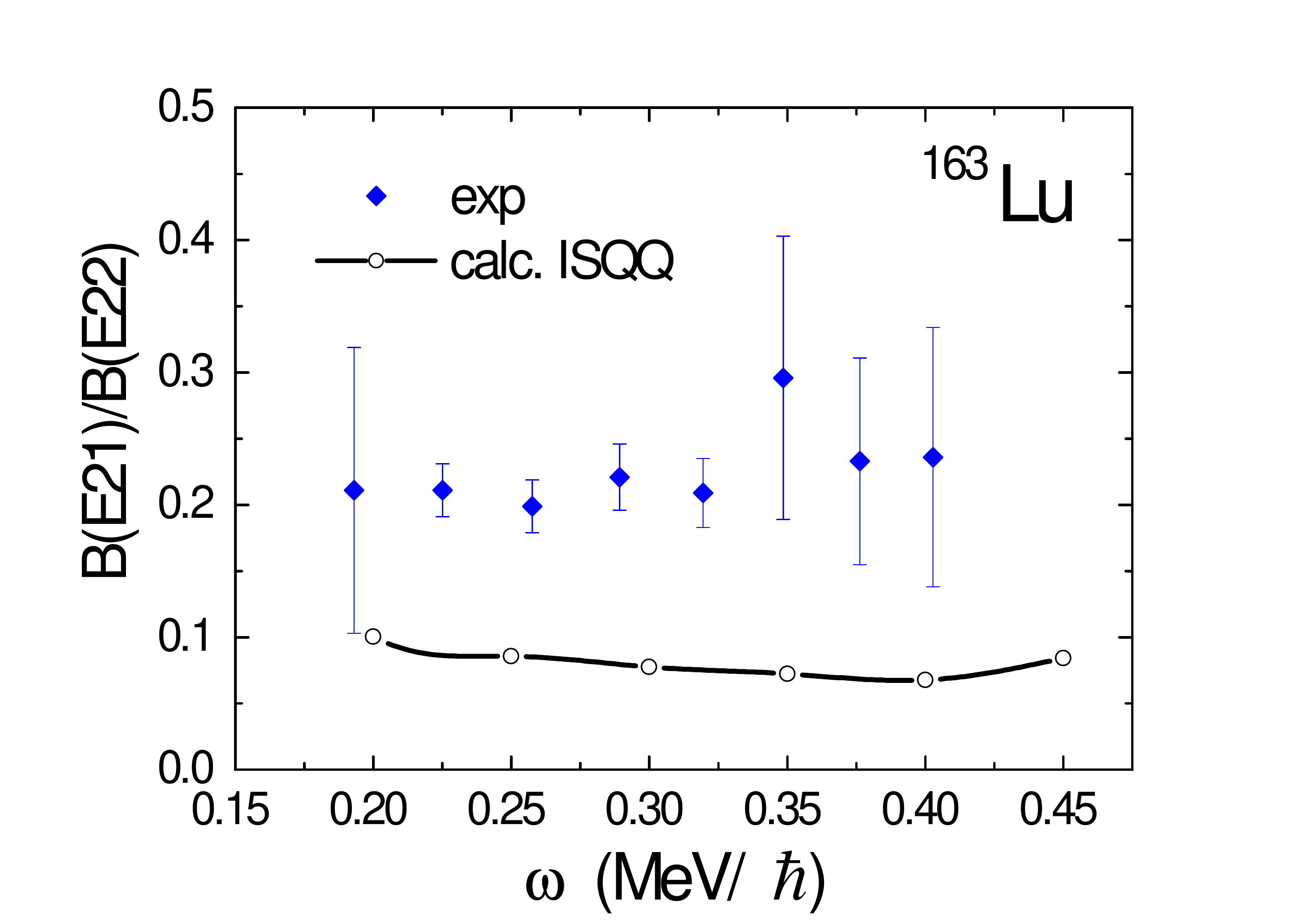} 
\caption{(Color online)  The  ratios B(E21)/B(E22) between the inter  and and intra band reduced  
transition probabilities B(E2, $I\rightarrow I-1$)$_{out}$/ B(E2, $I\rightarrow I-2$)$_{in}$
 for the transitions between the TSD wobbling band 
and the TSD ground band in  $^{163}$Lu. Notations as in Fig.\,\ref{ewoba}.}
 \label{BE21QQa} 
\end{figure}

\begin{figure}[htbp] 
\vspace*{-3cm}\includegraphics[clip,width=13cm]{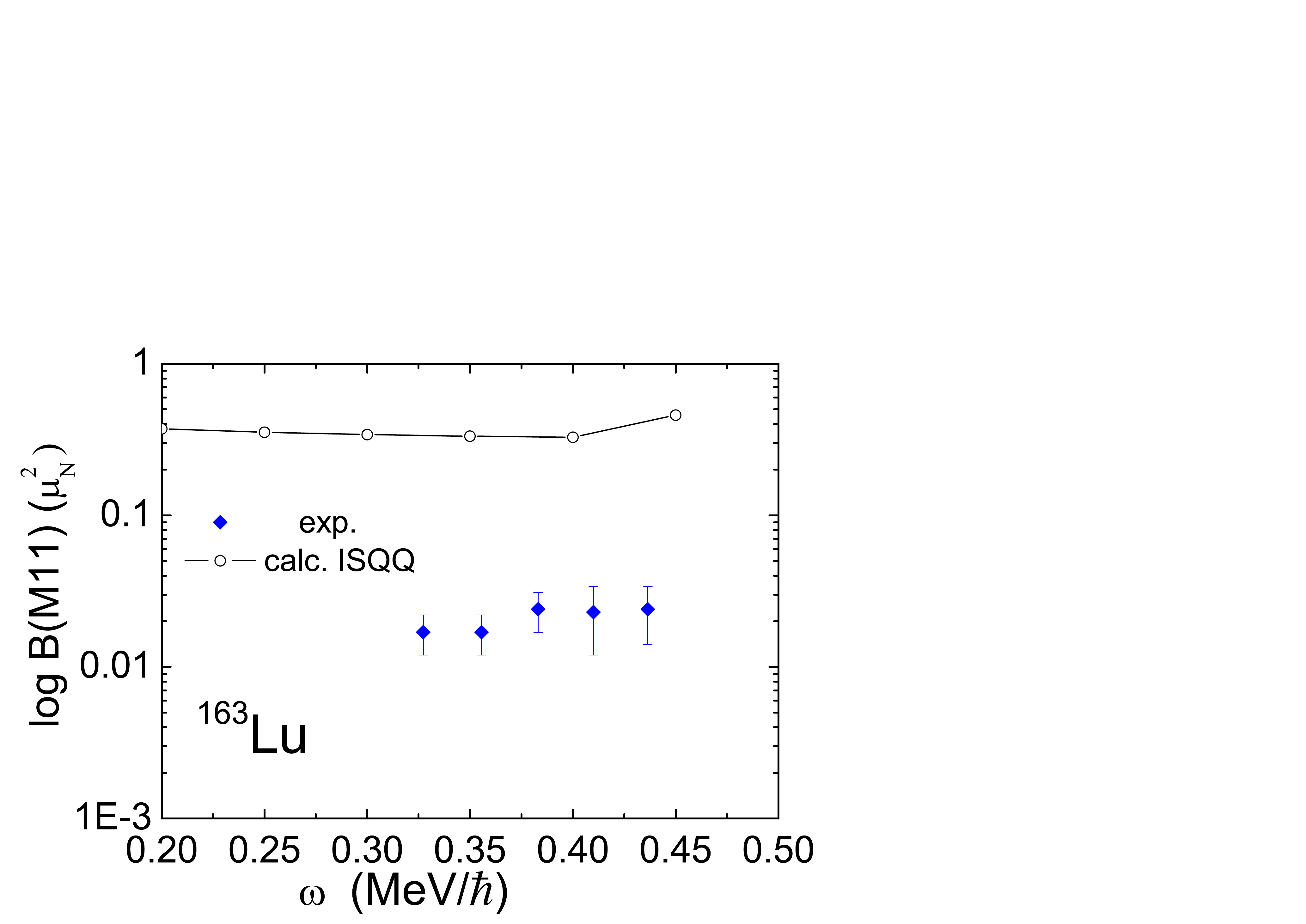} 
  \caption{(Color online)   The reduced transition probabilities B(M11)= B(M1, $(I\rightarrow I-1)$)$_{out}$  for the   transitions between the TSD wobbling band 
to the TSD ground band in  $^{163}$Lu. Notations as in Fig.\,\ref{ewoba}.}
 \label{BM11QQa} 
\end{figure}

In Figs.\,(\ref{ewoba} - \ref{BM11QQa}) we present the 
QRPA results for the wobbling energies and the inter band B$_{out}$(E2, $I\rightarrow I-1$) and B$_{out}$(M1, $I\rightarrow I-1$) values.
 The reduced transition probabilities of the upward transitions $I\rightarrow I+1$ are  
at least one order smaller and not displayed. The calculated wobbling energies 
$E_{_{QRPA}}(\omega)$ follow the decreasing tendency 
of the measured ones, which is characteristic for transverse wobbling. However, 
 they are substantially below the experiment. At $\omega$=0.45 MeV  the frequency becomes zero, which  
signalizes the change  to a permanent  tilt of the rotational  axis  away from the short axis. 
The experimental  wobbling energies
decrease  linearly  up to $\omega$=0.60 MeV.  The calculated ratios between the inter  and and intra band   
transition probabilities B(E2)$_{out}$/B(E2)$_{in}$ =B(E2, $I\rightarrow I-1$)/  B(E2, $I\rightarrow I-2$) 
reach only one  half of the measured values, whereas the calculated B(M1, $I\rightarrow I-1$) exceed the experimental ones by  a factor ten. 
Our results are similar to the ones of Ref. \cite{Ma02}, who used the QRPA version for ISQQ interaction in  the body fixed frame.
The deviations from experiment are about the same.


\subsection{ QRPA  for freely chosen shape parameters}\label{sec:ISQQsr}
The wobbling mode is sensitive to the ratios between the three moments of inertia, which strongly change
with the triaxality parameter $\gamma$.  The ISQQ coupling constant $\kappa_\circ=0.01960$ used in the preceding
section was adjusted to obtain a mean field deformation of $\varepsilon=0.4$.   
The sc values of $\gamma \approx 10^\circ$  obtained with the coupling constant 
fixed this way are substantially smaller than the values  that are 
calculated by minimizing the Nilsson-Strutinsky energy functional, which are given in Tab. \ref{tab:strutdefs}. 
Ref. \cite{Sh08} demonstrated  that larger values of  $\gamma$ increase the ratio $B(E2)_{out}/B(E2)_{in}$ between
the inter and intra band transitions. Their QPPR version in the body fixed frame does not use the 
selfconsistency in an explicit way, allowing them to freely choose the deformation of the mean field.  
In order to investigate this possibility we need give up the 
selfconsistency requirement, Eq.\,(\ref{scepsga}), 
between the shape parameters and the QQ interaction in Eq. (\ref{H2}) with the common strength parameter $\kappa_{_0}$.
This means we use the same deformed mean field Hamiltonian $\hat h$, Eq.\,(\ref{htac})
as before but the values of $(\varepsilon,\gamma)$ shall be at our disposal. 
Selfconsistency is only locally restored by 
constructing the residual interaction from the requirement that the resulting 
Hamiltonian $\hat H=\hat h+{V}_{4qp}$ becomes rotational invariant. Such "symmetry-restoring interaction"   
 \cite{Ba75,Faess}   
\begin{equation}
\label{Hfull}
{V}_{4qp} = -\frac{1}{2}\sum_{m=1}^3 \kappa _m F_m^2 
\end{equation}
is built from the squares of the commutators of the quasiparticle  Hamiltonian $\hat h$  and the angular momentum components $J_{m=1,2,3}$: 
 \begin{equation}
\label{force}
    F_m = [\hat h, i J_m].
\end{equation}
The strength constants $\kappa _m$ are determined by demanding rotational invariance via the commutator 
\begin{equation}
\label{commut}
[\hat H,i J_m] = 
[h-\frac{1}{2}\sum_{n=1}^3 \kappa _{n} F_{n}^2 , iJ_m] = 0 
\end{equation}
which can be satisfied on average $\langle[\hat H,i J_m]\rangle=0$ by fixing the strength constants according to
\begin{equation}
\label{kappa}
\kappa_m^{-1} = \langle [[\hat h,iJ_m],iJ_m]\rangle
\end{equation}
where $|\rangle $ is  the reference quasiparticle configuration.

 This method can be applied to any mean field Hamiltonian $\hat h$, as  for instance in Ref.\,\cite{Shoji} to a deformed Woods-Saxon potential.
In our case the commutator (\ref{force}) with a  
quadrupole deformed field  generates again quadrupole operators. 
Now we evaluate the  commutators relations (\ref{force}, \ref{kappa}) 
explicitly. The results are written in terms
of the combined 
quadrupole operators $Q_{_{1\pm}}$ and $Q_{_{2\pm}}$ defined by
\begin{eqnarray}
\label{combin}
   Q_{_{1+}} &=& \frac{ Q_{_{1}}+  Q_{_{-1}}}{i\sqrt{2}},
\quad Q_{_{1-}} = \frac{ Q_{_{1}}-  Q_{_{-1}}}{\sqrt{2}},\nonumber\\
    Q_{_{2+}} &=& \frac{  Q_{_2}+  Q_{_{-2}}}{\sqrt{2}},\quad
   Q_{_{2-}} = \frac{  Q_{_2}-  Q_{_{-2}}}{i\sqrt{2}}.
\end{eqnarray}
Introducing for given deformations $(\varepsilon,\gamma)$ the constants ${\cal Q }$ and $\tilde\gamma$ defined by
\begin{eqnarray}
\label{Qangle}
{\cal Q } = \sqrt{\langle Q_0\rangle ^2 + \langle Q_{2+}\rangle ^2}\,,\quad\quad \nonumber\\ \quad\quad
\sin{\tilde\gamma} =-\frac{\langle Q_{2+}\rangle}{\cal Q}, \quad \quad
\cos{\tilde\gamma} =\frac{\langle Q_{0}\rangle}{\cal Q} 
\end{eqnarray}
the  Hamiltonian\,(\ref{eq:H1}) with the interaction\,(\ref{Hfull}) takes the form 
\begin{eqnarray}
\label{Hfull3}
\hat H = \hat h +\frac{1}{3} \frac{\hbar\omega_\circ \varepsilon}{\cal Q}\,\, [\,\,
\frac{\sin{\gamma}}{\,\sin{\tilde\gamma}}\,Q_{_{2-}}^{\,2} +
\frac{\sin{(\gamma+2\pi/3)}}{\,\sin{(\tilde\gamma+2\pi/3})}\,Q_{_{1+}}^{\,2} +\nonumber\\+\,
\frac{\sin{(\gamma-2\pi/3)}}{\,\sin{(\tilde\gamma-2\pi/3})}\,Q_{_{1-}}^{\,2} \,\, ] \quad \quad\quad \quad\quad,
\end{eqnarray}
where the constants $\kappa _m$ are expressed in terms of  ${\cal Q }$ and $\tilde\gamma$.
Hence, the residual interaction of the Hamiltonian $\hat H$, needed  for the QRPA, is fully determined by 
the mean field part $\hat h$ , in our case   by its deformation parameters $(\varepsilon,\gamma)$.

The "symmetry-restoring interaction"  includes the selfconsistent treatment of the Hamiltonian (\ref{H2}) as
 a special case.  Using the notation 
(\ref{combin})
\begin{equation}
\hat H = \hat h 
-\frac{\kappa_\circ}{2} \sum_{\mu=0,1\pm,2\pm} Q_{\mu} ^2
\end{equation}
which in comparison to the one in Eq.(\ref{Hfull3}) contains additionally the terms 
$\,Q_{_{0}}^{\,2} $ and $\,Q_{_{2+}}^{\,2} $  that are driving the beta-gamma vibrations.
In this case one has to search for deformations $(\varepsilon, \gamma)_{sc}$
which comply with the selfconsistent conditions (cf. Eq.\,(\ref{scepsga}))
\begin{eqnarray}
\label{selfco2}
 \frac{\kappa_\circ}{2}&=& \frac{1}{3}\frac{\hbar\omega_\circ \varepsilon}{\cal Q}\,\,,\quad\quad 
\sin{\gamma}=-\frac{\langle Q_{2+}\rangle}{\cal Q} = \,\sin{\tilde\gamma}.
 \end{eqnarray}
 According to Eq.\,(\ref{Hfull3}) one recognizes that for the selfconsistent deformation 
$(\varepsilon,\gamma)_{sc}$
 the common prefactor in the Hamiltonian (\ref{Hfull3}) becomes equal to 
$\kappa_\circ/2$ and the three ratios of the Sin terms  become one. 
Thus, for the sc deformations the  Hamiltonian $\hat H$, Eq.\,(\ref{H2}) is fully rotational invariant, 
i.e. the commutator relations (\ref{commut}) are exactly satisfied.

At variance with the standard $QQ$-Hamiltonian (\ref{H2}),  the  coupling strengths of the three  interaction terms $Q_{k\pm} ^2$ in Eq. (\ref{Hfull3}) 
are  not equal for arbitrary choice of the deformation parameters $(\varepsilon,\gamma)$. 
With the values of  $\kappa_{1,2,3}$ obtained from Eq.(\ref{kappa}) rotational symmetry is achieved  $locally$ because the
commutator relations (\ref{commut}) are satisfied on average.
This is in accordance with fact that  the QRPA  treats  
the wobbling motion as a small angle vibration. Local rotational invariance ensure that  the spurious rotational excitations can 
be removed as the ones with the energies $E_{_{QRPA}}=0$ and $\hbar\omega$ (as in the sc case).

Below we present the results of a QRPA calculation with the Hamiltonian $\hat H$ of Eq.(\ref{Hfull3}) using   
the deformation parameters $(\varepsilon,\gamma)$ of Tab.\,\ref{tab:strutdefs}, which were found by a 
Nilsson-Strutinsky minimization \cite{Go04}.
\begin{table}[t]  
\caption{   
Deformation parameters $(\varepsilon, \gamma)$ of $^{163}$Lu in the frequency region $\omega=$ 0.15-0.55 MeV/$\hbar$ obtained from a 
Nilsson-Strutinsky (NS) minimization \cite{Go04}} 
\label{tab:strutdefs}
\begin{ruledtabular} 
\begin{tabular}{ccc} 
$\omega$(MeV/$\hbar$) & $\varepsilon$& $\gamma$(deg) \\\hline
0.15&	0.3815&	18.75\\	
0.2&	0.3892&	19.2\\	
0.25&	0.3968&	19.64\\	
0.3&	0.4044&	20.12\\
0.35&	0.408&	20.41\\
0.4&	0.3991&	20.72\\	
0.45&	0.3908&	21.3\\
0.5&	0.3852&	21.78\\
0.55&	0.3812&	22.34
\end{tabular} 
\end{ruledtabular} 
\end{table} 
Notice that $\gamma$ is about$^\circ$ larger than the corresponding selfconsistent values. 
\begin{figure}[htbp] 
\includegraphics[clip,width=\linewidth]{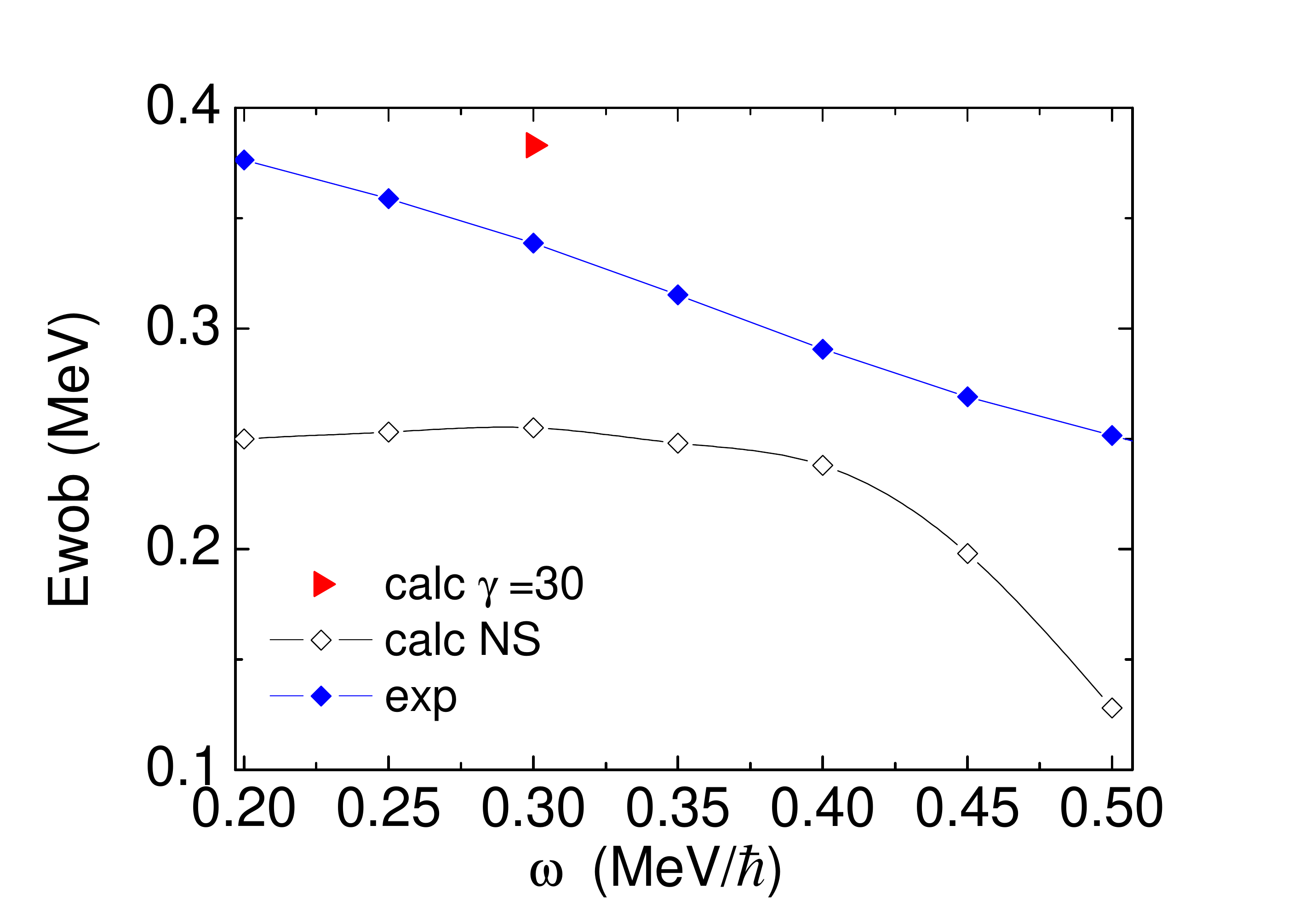} 
  \caption{(Color online)  Wobbling frequencies in $^{163}$Lu as a function of the 
rotational frequency. Experimental values (blue diamonds) are from \cite{Go04}. 
The calculated values (solid line) are obtained with the Nilsson-Strutinsky (NS) deformations  in table\,\ref{tab:strutdefs}. 
The single  value (red triangle) is found 
for the deformation point ($\epsilon=0.4,\gamma=30$ deg).} 
  \label{ewoba-NS} 
\end{figure} 

Fig.\,\ref{ewoba-NS} shows  the calculated wobbling frequencies together with the experimental values. 
Compared to the wobbling frequencies of the sc ISQQ model (cf.\,Fig.\,\ref{ewoba}) the calculation with the Nilsson-Strutinsky deformations gives a
flatter $\omega$ dependence, and the break down of the QRPA is slightly retarded. As seen in Fig. 15 of  our study \cite{FD14},
 the QRPA wobbling frequency curve resembles the one
obtained by applying the HFA approximation to the Quasiparticle Triaxial Rotor QTR)description of transverse wobbling in $^{163}$Lu using microscopic
moments of inertia calculated by means of the TAC model. The HFA is a small-amplitude approximation like QRPA. The full quantal solution
of the    QTR shows a gradual decrease of the wobbling frequency with frequency, which is closer to experiment (cf. Fig. 15 of \cite{FD14}).  

\begin{figure}[htbp] 
\includegraphics[clip,width=\linewidth]{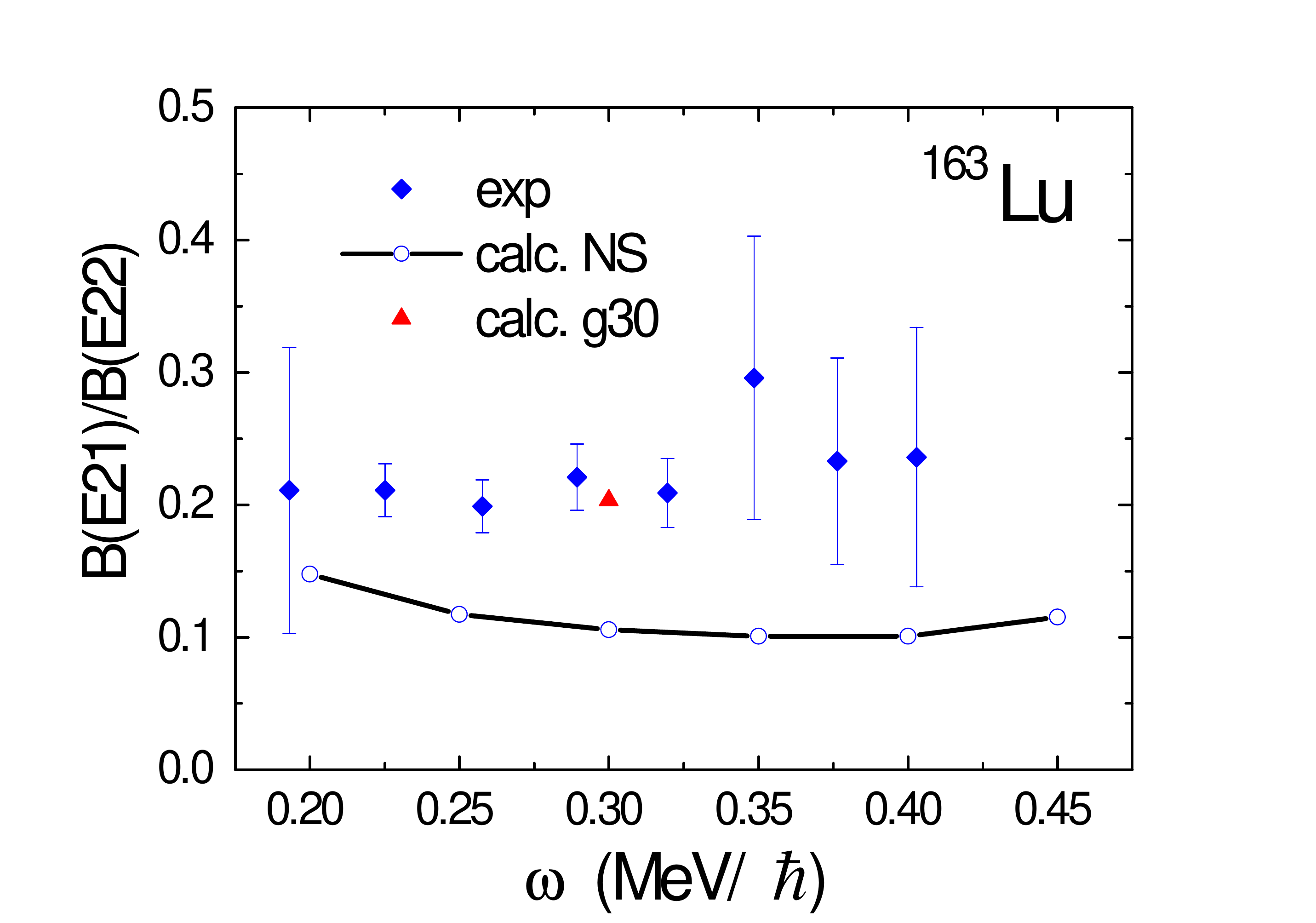} 
 \caption{(Color online)   The  ratios B(E21)/B(E22) between the inter  and and intra band reduced  
transition probabilities B(E2, $I\rightarrow I-1$)$_{out}$/ B(E2, $I\rightarrow I-2$)$_{in}$
 for the transitions between the TSD wobbling band 
and the TSD ground band in  $^{163}$Lu. 
Notations as in Fig.\,\ref{ewoba-NS}.}
 \label{BE21-NS} 
\end{figure}

\begin{figure}[htbp] 
\includegraphics[clip,width=10cm]{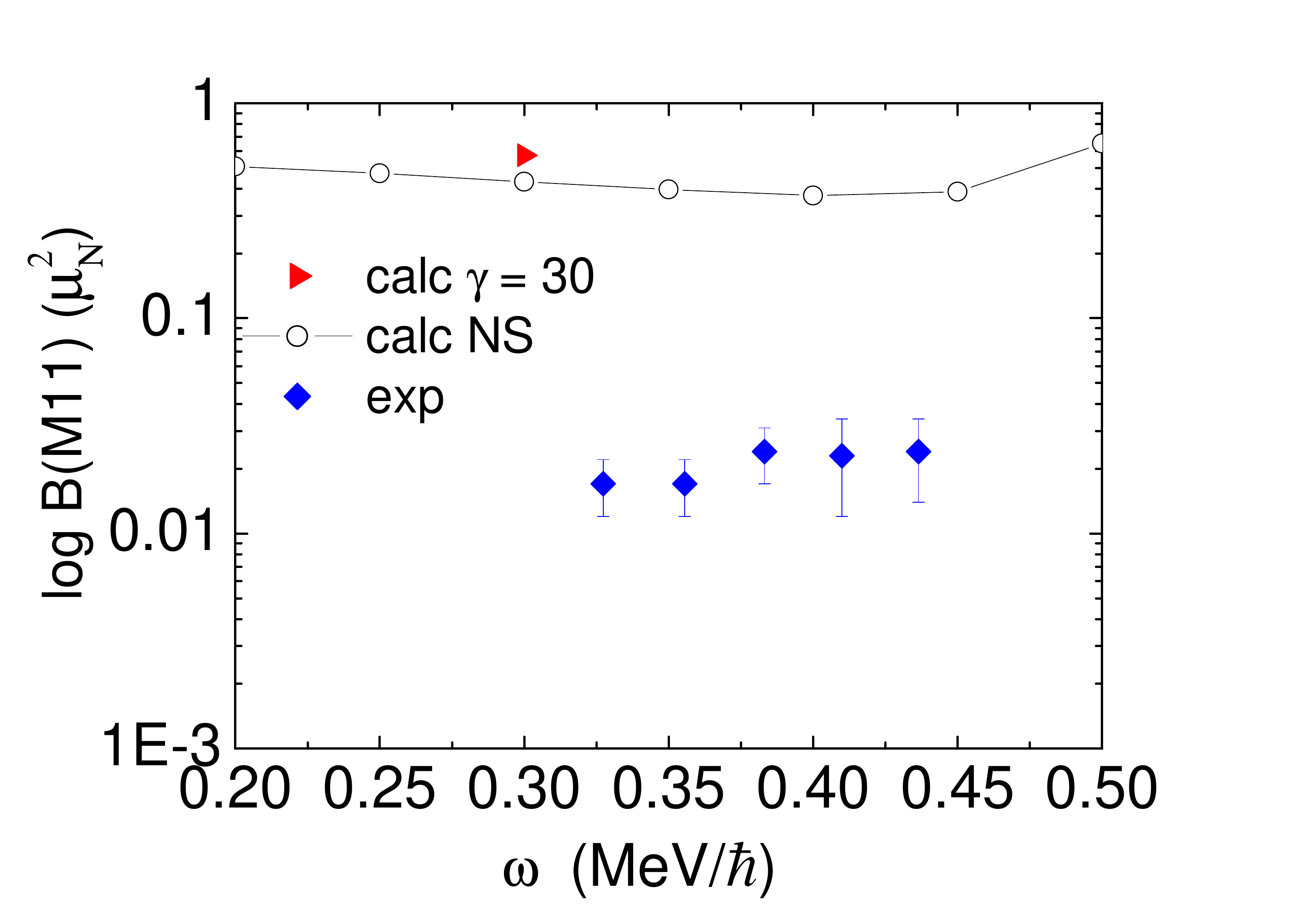} 
  \caption{(Color online)    The reduced transition probabilities B(M11)= B(M1, $(I\rightarrow I-1)$)$_{out}$  for the   transitions between the TSD wobbling band 
to the TSD ground band in  $^{163}$Lu. Notations as in Fig.\,\ref{ewoba-NS}.}
 \label{BM11-NS} 
\end{figure}

Comparing Fig.\,\ref{BE21QQa} with Fig.\,\ref{BE21-NS}. shows that the larger $\gamma$ values lead to a 20 \% increase of the ratio B(E2)$_{out}$/B(E2)$_{in}$.
No reduction is obtained for the magnetic inter band  transition strength as seen comparing Fig.\,\ref{BM11QQa} and Fig.\,\ref{BM11-NS}. Hence with the 
larger $\gamma$ values predicted by the Nilsson-Strutinsky calculation 
and the symmetry restoring QRPA we only accomplish a marginally better description of the TSD band properties. 


We tried the case of maximal triaxiality 
$\gamma=30^\circ$ and $\varepsilon=0.4$ for $\hbar \omega$=0.3 MeV. The results are included in Figs.(\ref{ewoba-NS}-\ref{BM11-NS}). 
The wobbling frequency  is enlarged, and even exceeds the experimental value. The ratio 
B(E2)$_{out}$/B(E2)$_{in}$  is about right, such that it could be adjusted by choosing  an
appropriate $\gamma$ value between 25$^\circ$ and $30^\circ$. 
However, the small B(M1)$_{out}$ values remain unexplained. On the other hand,  Ref. \cite{Shoji} demonstrated  
that QRPA  based on the Woods-Saxon potential and the pertinent symmetry restoring interaction
 gives the experimental B(E2)$_{out}$/B(E2)$_{in}$ ratio for $\gamma \approx 20^\circ$. In Ref. \cite{Sh08} the authors relate 
 the small values of the ratio to artifacts of the Nilsson potential, which do not exist for the Woods-Saxon potential.
 In view of this, we attribute the small ratio  of B(E2)$_{out}$/B(E2)$_{in}$  obtained in our calculations to our choice  of the 
 Nilsson potential. Ref.    \cite{Shoji} finds a comparable low wobbling frequency as we do and does not present the B(M1)$_{out}$ values.
 Thus, at this point we conclude 
    the use of the isoscalar QQ residual interaction (including its symmetry restoring variants)   only partly explains the experimental findings. 
 This is the reason for study additional residual interactions  in the following.

\begin{figure}
\includegraphics[clip,width=\linewidth]{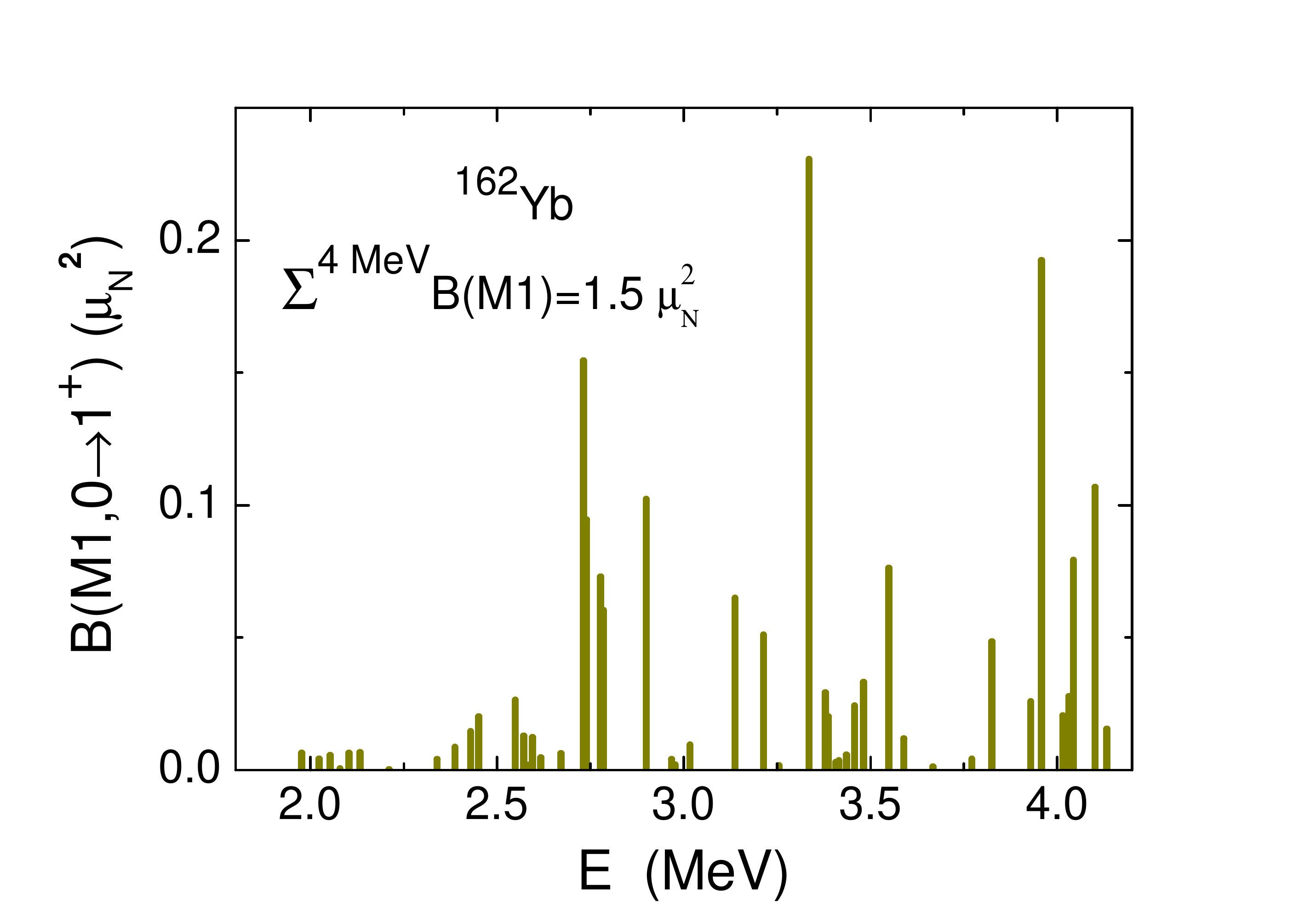} 
\caption{(Color online) B(M1) distribution of $^{162}$Yb obtained by rpa with the LL interaction, Eq.\,(\ref{VLL}) choosing the strength constant $\kappa_{_{LL}}=0.5$\,MeV$/\hbar^2$. 
The fragmented B(M1) strength adds up from 0 to 4 MeV to a sum strength of 1.5 $\mu_N^2$
that can be interpreted as scissors strength. }
 \label{m1yb} 
\end{figure}

  \section{ Additional residual interaction terms}\label{sec:AddRI}
  
  The experimental fact that the inter band M1 transitions of the wobbling mode
are strongly suppressed in comparison to the inter band  E2 transitions 
is a major motivation to study further interaction terms aside the QQ interaction 
considered so far. It is the question, what makes that the magnetic de-excitation so small.
It is known that the scissors mode collects the low-lying M1 strength which is  concentrated higher up in the energy region of 3-4 MeV \cite{Heyde}. 
A possible mechanism for suppressing the M1 strength of low-energy states is shifting it to the  scissors mode,  like 
the electric dipole strength of low energy states is shifted to the Giant Dipole Resonance.

   Before presenting the results of our QRPA calculations 
with additional interaction terms a note about the removal the spurious rotational modes  is in order. 
 When adding  interaction terms the strength constants of which are not fixed by selfconsistency  or rotational invariance the 
rotational modes  shift away from their true energies $E_{_{QRPA}}=0,\,\hbar\omega$ and mix with the wobbling mode,
such that the results are distorted by spurious effects. Therefore 
we apply  the method proposed in Ref.~\cite{FD05} to eliminate  the spurious modes. The QRPA Hamiltonian is 
complemented by the IS term  $\kappa_j {\bf J} \cdot {\bf J}$ which acts like a spring force for the unwanted angle vibrations 
of the total angular momentum ${\bf J}$ in the rotating system.
Choosing the stiffness parameter large, as $\kappa_j \ge 10^2$,  the 
excitation energies for the rotational spurious states are shifted far outside the considered energy range, which prevents them from mixing with the physical modes.

Our first modification was motivated by the purely collective picture of  the scissors mode being an 
angle vibration of the proton system against the neutron system with an IV QQ restoring force  \cite{LoJudice}. 
Accordingly,  we added to the ISQQ Hamiltonian (\ref{H2}) an  IV QQ interaction term built from the operators $\hat Q_{m}^{iv} = \hat Q_{m}(\pi)$ - $\hat Q_{m}(\nu)$. 
Knowing the selfconsistent strength  $\kappa_\circ$ from table~\ref{tab:defs} we set the isovector strength $\kappa_\circ^{iv}=r \,\kappa_\circ$ where the value of the ratio $r$
was varied in the range -1.5 to -\,3.5 \cite{Bo75}. This addition lead to only a minor  change of the B(E2/M1) transition probabilities. However, it increased the wobbling frequency,
 such that  the experimental wobbling frequencies could be fitted  by choosing an appropriate value of $r$.

Second, we considered the spin-spin (SS) interaction, because it
has been successfully applied in connection with the scissors mode to explain the systematic accumulation of 1$^+$ states 
between 3-5 MeV with considerable M1 decay strengths \cite{Faess}.
 We  included both the IS and the isovector (IV) spin-spin interactions defined by
\begin{eqnarray} \label{SS}
V^{(is,iv)}_{LL}&=&\sum_{_{m=-1,1}}(-1)^m\,{\hat{S}^{(is,iv)}_{_m}\hat{S}^{(is,iv)}_{_{-m}}}, \nonumber\\
\hat{S}^{(is,iv)}_{_m}&=& \hat S_{m}(\tau =+1)\pm\hat S_{m}(\tau =-1).
\end{eqnarray}
We determined the SS strength parameters  by
extrapolating the A-dependent strength parameters given in the work of De Coster and Heyde \cite{deCoster} used there for QRPA calculations of the 1$^+$.
 The SS interactions are then added to the selfconsistent ISQQ Hamiltonian (\ref{H2}) described in Sec. \ref{sec:ISQQsc}.
The results of the QRPA calculation  for the frequency $\hbar\omega=0.3$ MeV/$\hbar$  can be summarized as follows: The 
IS and IV SS terms have only negligible effects on both the wobbling energy and the B(E2/M1) transition probabilities. 
The lowering of the B(M1)$_{out}$ value is small,  i.e. there is not much  shift the M1 strength into the scissor region.

Our third modification was motivated by the interpretation that
the scissors mode represents an angle vibration of the orbital a.m.\,vector $\vec {\bf L}_\pi$ of the protons  versus the orbital a.m.\,vector $\vec {\bf L}_\nu$ of the neutrons. 
Accordingly, we complemented the ISQQ Hamiltonian (\ref{H2}) by an interaction term that is composed of the isovector orbital angular momenta:        
\begin{equation}
\label{VLL}
V_{_{LL}}=\kappa_{_{LL}}(\vec {\bf L}_\pi-\vec {\bf L}_\nu )^2.
\end{equation} 
Ref. \cite{Rusev06} successfully used an interaction of the type  $V_{JJ}=\kappa_{_{JJ}}(\vec {\bf J}_\pi-\vec {\bf J}_\nu )^2$ to describe the
M1 strength in the scissors region of the Mo isotopes. We checked  that such  IV JJ interaction gives nearly the same results  as the LL interaction
when the coupling constant is appropriately chosen.

\begin{figure}[htbp] 
\vspace*{0cm}\includegraphics[clip,width=\linewidth]{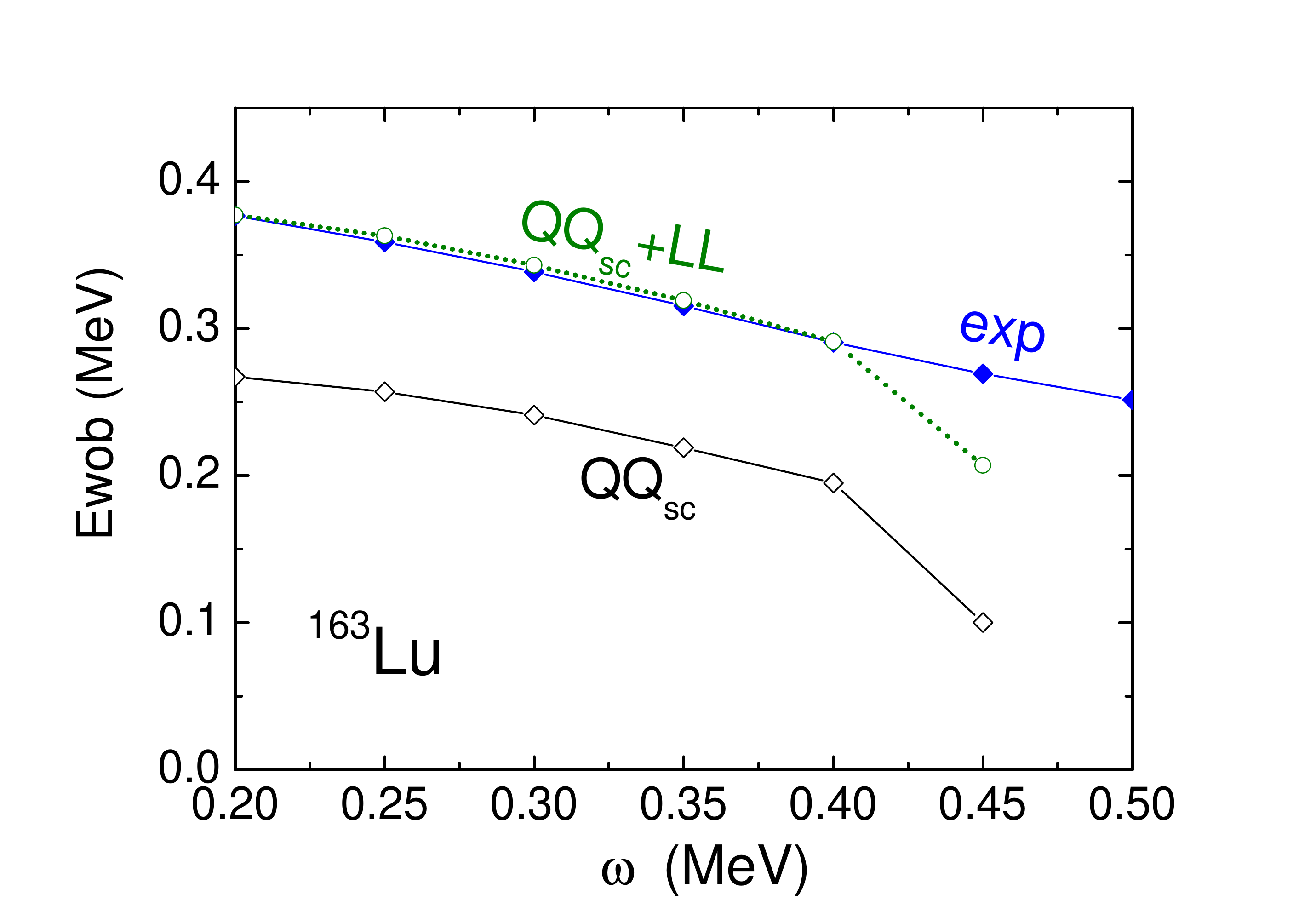}  
  \caption{(Color online)  Wobbling frequencies in $^{163}$Lu as a function of the 
rotational frequency. Experimental values (blue diamonds) are from \cite{Go04}. Calculated values  are obtained with QRPA using  
(solid line) selfconsistent ISQQ interaction and (green dotted line) additionally LL interaction
(see Eq.\,(\ref{VLL}))} 
  \label{Ewob-LL} 
\end{figure} 

\begin{figure}[t] 
\includegraphics[clip,width=\linewidth]{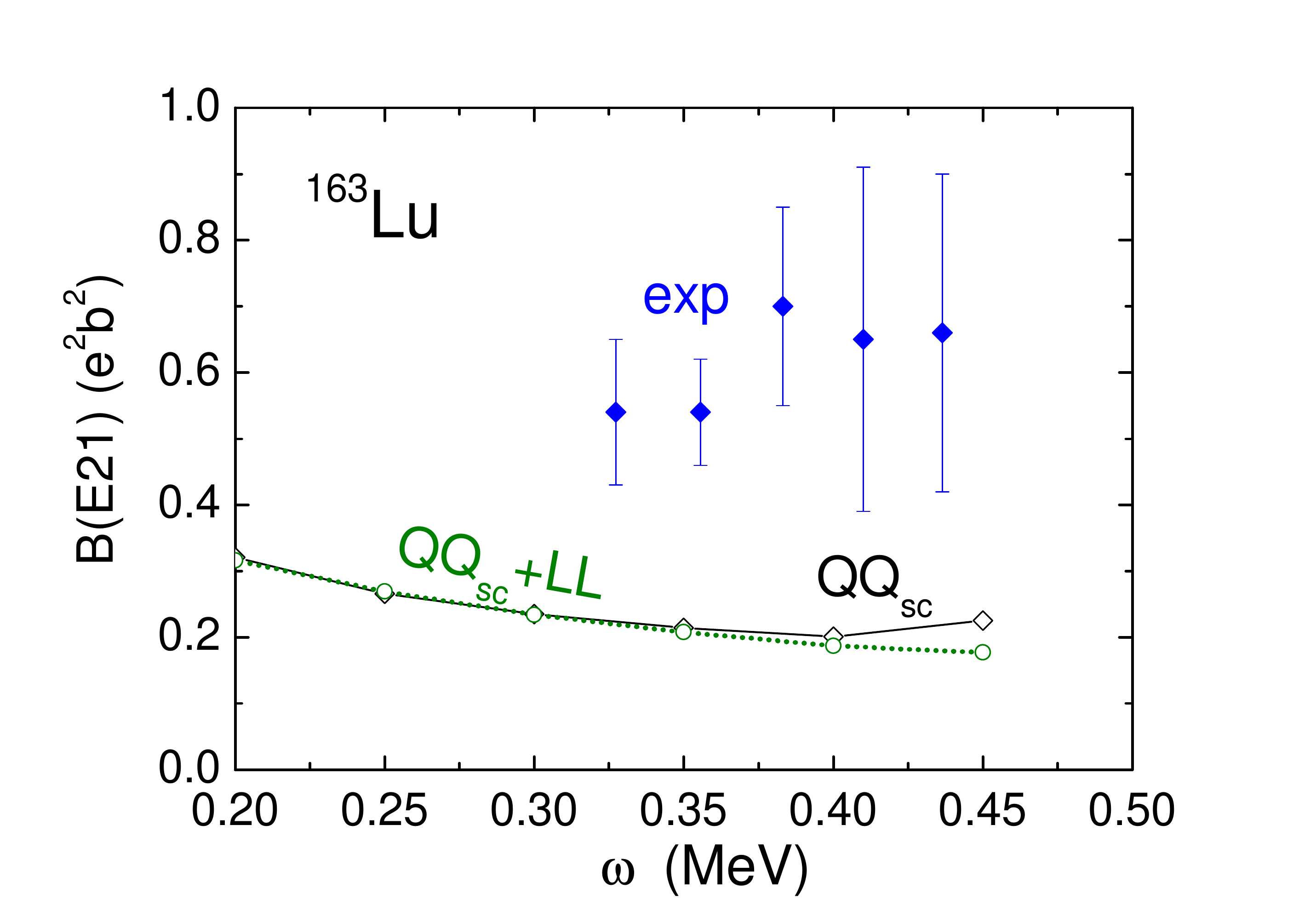}  
 \caption{(Color online)   The  reduced  
transition probabilities B(E21)=B(E2, $I\rightarrow I-1$)$_{out}$
 for the transitions between the TSD wobbling band 
and the TSD ground band in  $^{163}$Lu. 
Calculated values  are obtained with QRPA using  
(solid line) selfconsistent ISQQ interaction and (green dotted line) additionally LL interaction
(see Eq.\,(\ref{VLL})).}
 \label{BE21-LL} 
\end{figure}

\begin{figure}[t] 
\includegraphics[clip,width=\linewidth]{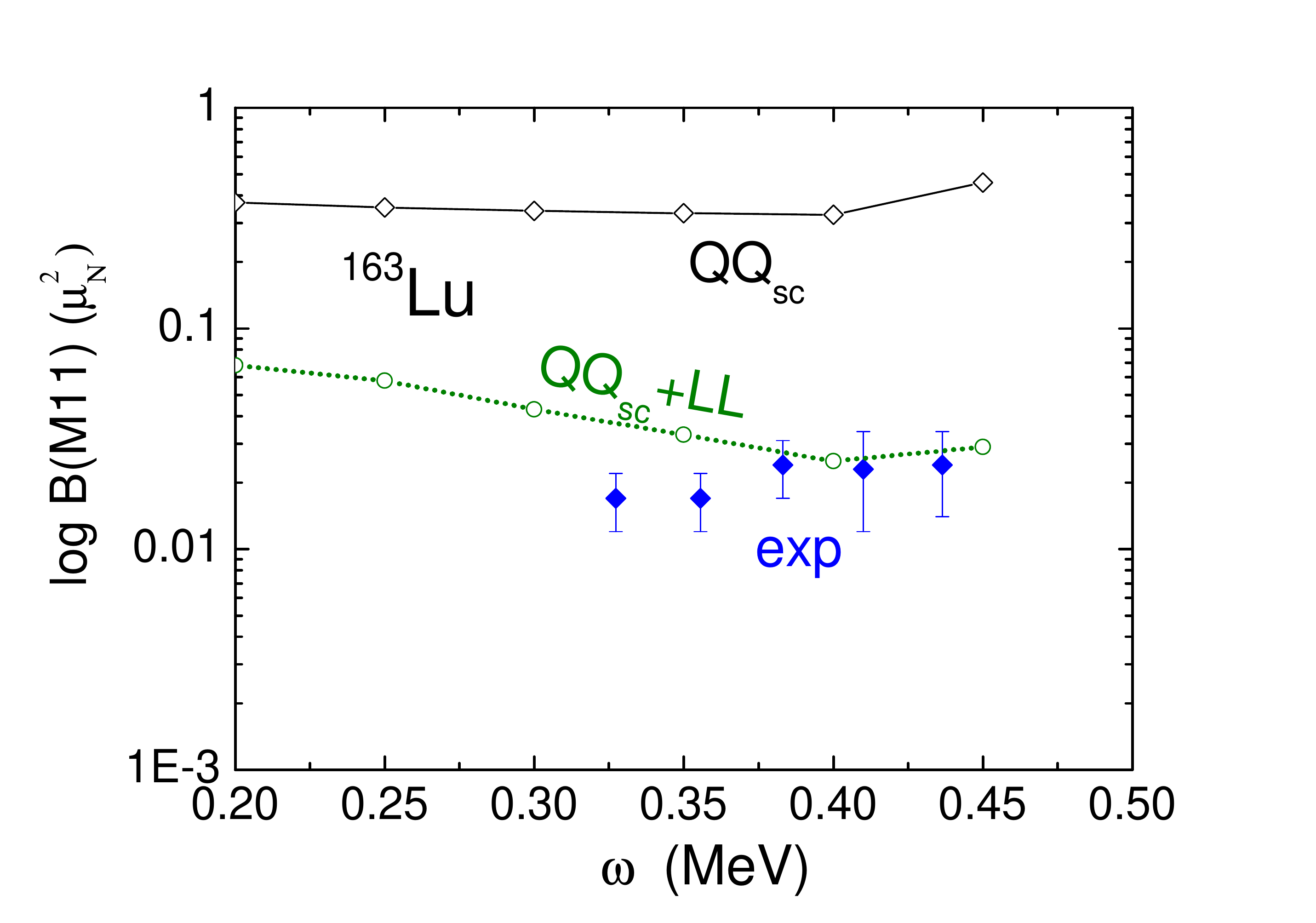}
\caption{(Color online)  The reduced transition probabilities B(M11)= B(M1, $(I\rightarrow I-1)$)$_{out}$  for the   transitions between the TSD wobbling band 
to the TSD ground band in  $^{163}$Lu.
Calculated values  are obtained with QRPA using  
(blue line) selfconsistent ISQQ interaction and (green dotted line) additionally LL interaction
(see Eq.\,(\ref{VLL})).}
 \label{BM11-LL} 
\end{figure}

The effects of adding the LL interaction are shown in Figs.(\ref{Ewob-LL}-
\ref{BM11-LL}). The calculated wobbling energy increases due to the repulsive 
LL term. We find a good match to the experimental curve when choosing the strength constant $\kappa_{LL}=0.5$ MeV/$\hbar^2$. The inter band
E2 transitions stay almost unchanged, which is expected from a current-current interaction. The  same value of $\kappa_{LL}$  gives the 
desired suppression of the B(M1) transition strength, which comes close to the measured values. 

Hence, the QRPA with additional LL interaction  is capable of providing a satisfactory description
of the wobbling frequencies and of the magnetic properties. This raises the question whether the adjusted 
coupling parameter $\kappa_{LL}$ is consistent with the experimental information 
about the scissors mode built on the ground states of the even-even neighbors.  We calculated the distribution of B(M1,\,0\,$\to$\,$1^+$) from the ground state of $^{162}$Yb 
using the same QRPA approach as for the wobbling mode in $^{163}$Lu. The deformation
$\beta=0.225$  from was taken Ref.\,\cite{ Moeller}, and  the value  
$\kappa_{_{LL}}$$=$\,0.5\,MeV/$\hbar^2$ used for the LL interaction. The resulting distribution  is shown in Fig.\,\ref{m1yb} for interval E=2-4\,MeV,
 which is the suggested region of the scissors mode. There is no experimental information for the unstable nuclid $^{162}$Yb about the distribution of 1$^+$ states to compare with.  
However, the systematics of the summed B(M1) strength presented in Refs.\,\cite{Ziegler,Heyde} provides a clue concerning the 
coupling constant.   Our   value  $\kappa_{_{LL}}$$=$\,0.5\,MeV/$\hbar^2$ gives a summed strength $\Sigma B(M1)$$\,\approx$\,1.5\,$\mu_N^2$ for the 1$^+$ excitations between 0-4 MeV,
 which agrees with the value from the systematics for the deformation $\beta=0.225$ of $^{162}$Yb.  The agreement indicates that the coupling of the transverse wobbling to the scissors mode
 at high spin and the M1 strength of the low-spin scissors mode can be accounted for by one and the same value of $\kappa_{LL}$.

 The improvements achieved by including   the IV LL interaction term can be taken as an indication that the wobbling motion is 
 not a pure orientation vibration of  the quadrupole mass tensor  with respect to the angular momentum  vector.  It implies a
 coupling to  vibrations of the proton and neutron currents against each other (see the interpretation of the scissors mode in Ref. \cite{Aberg84}). 
The microscopical origin of such schematic interaction of the current-current type remains obscure at this point. However,
it is noted that Ref. \cite{Faess} well describe the low-spin scissors mode in the framework of QRPA based on a symmetry restoring 
interaction that is derived from a deformed Woods-Saxon potential. The pertinent commutators with the deformed spin-orbit potential 
will generate IV terms of containing the the momentum and spin operators, which may be schematically accounted for by the LL interaction.
It would be interesting to see how QRPA based on the symmetry restoring interaction of Ref. \cite{Faess}   describes transverse wobbling.


 \section{ Summary and Conclusions}\label{sec:SC}
 
 The transverse wobbling mode in $^{163}$Lu has been reinvestigated in the frame work of Quasiparticle Random Phase Approximation.
 The QRPA calculations were based on the rotating mean field that consisted of a deformed Nilsson potential and an attenuated monopole
 pair field. Various versions of the residual interaction were investigated.  For all variants studied, The QRPA wobbling frequencies 
 decrease with the rotational frequency, so confirming the transfers character of the solution.  
The results obtained with an isoscalar Quadrupole-Quadrupole interaction and 
 selfconsistent deformation parameters in essence agree with previous QRPA calculations \cite{Ma02}, which used the same mean field
 Hamiltonian but another way of finding the solutions. 	 The calculated wobbling frequencies show the right descent with the rotational frequency 
 but are only 60\% of the experimental excitation energy. The  B(E2)$_{out}$/B(E2)$_{in}$  ratios for the inter band transitions connecting  the wobbling with the ground band 
 and the intra band transitions show the characteristic
collective enhancement, but are low by about a factor two.  The B(M1)$_{out}$ values of these inter band transitions are a factor 10 too large compared with experiment. 

To check whether the small  
selfconsistent values of the triaxiality parameter $\gamma\approx 10^\circ$ are the cause for the deviations,  
we tried the QRPA variant  based on a factorized residual interaction that is derived from the mean field by requiring local rotational invariance, which  allows
one to chose freely the deformation parameters. Using $\gamma\approx 20^\circ$, which is the equilibrium value of the Nilsson-Strutinsky Routhian, slightly moves 
the results toward the experimental values, however the discrepancies remain as substantial as before. Increasing the triaxiality to $\gamma\approx 30^\circ$
brings the  B(E2)$_{out}$/B(E2)$_{in}$  ratio up to the experimential value and the wobbling frequency a bit above the experimental one, such that $\gamma$ values somewhat below
30$^\circ$ will lead to a match with experiment. However, we do not favor this possibility for the following reasons.  Cranked mean field calculations 
generally predict smaller triaxiality of $\gamma\approx 20^\circ$.  Ref. \cite{Shoji} demonstrated that using a Woods-Saxon potential with  
$\gamma\approx 20^\circ$ gives the experimental B(E2)$_{out}$/B(E2)$_{in}$  ratios. As discussed next,  $\gamma\approx 30^\circ$ would 
shift the wobbling frequencies too high when the additional residual LL interaction is taken into account. 
Rather we share the view of Ref. \cite{Sh08} that the low  B(E2)$_{out}$/B(E2)$_{in}$  ratios are an artifact of the Nilsson potential, which will be removed by replacing it by a more 
realistic potential. 

Our study confirms  previous 
findings  that QRPA based on an isoscalar QQ-type interaction gives too small wobbling frequencies and too large inter band  B(M1) values. These discrepancies are removed by 
including a repulsive   isovector current-current interaction of the schematic form  $ \kappa_{_{LL}}(\vec {\bf L}_\pi-\vec {\bf L}_\nu )^2$, where $ \vec {\bf L}$ is the total orbital angular momentum.
This LL interaction couples the wobbling mode to the scissors mode, which represents a concentration of orbital M1 strength  in the  region E=3-4 MeV above the yrast line. 
The B(M1)$_{out}$ values are reduced, because M1 strength is shifted into the scissors region, and the wobbling frequencies increase because the LL interaction 
is repulsive.  The same interaction strength  $ \kappa_{_{LL}}$ generates  the right upshift of the wobbling frequencies and the right 
suppression of the  B(M1)$_{out}$ values  toward the experimental values.   Moreover, using the same $ \kappa_{_{LL}}$ value, QRPA on the ground state of the neighbor $^{162}$Yb
reproduces the cumulative M1 strength below 4 MeV, known from experimental systematics.  

Altogether, QRPA based on the  combination of the  isoscalar QQ	and isovector LL interaction well reproduces the experiments on transverse wobbling of the 
triaxial strongly deformed	nuclide $^{163}$Lu. The mode represents mainly an oscillation of the triaxial charge distribution relative to the angular momentum vector, which
is manifest by strong E2 transitions from the one-phonon to the zero-phonon wobbling bands. Additionally,  it contains a substantial admixture of scissors-like oscillations of the 
proton currents against the neutron currents, which increase the wobbling frequency and reduce the M1 transition strength  between the wobbling
 bands by a factor of 10.  In the  other case of a well studied example
of transverse wobbling, $^{135}$Pr, the QTR calculations in Ref. \cite{135Pr}, which do not take into account the coupling to the scissors mode, overestimate the
B(M1)$_{out}$ values by a factor of three. One expects a weaker coupling to the scissors mode, because  $^{135}$Pr is much less deformed than $^{163}$Lu and it is known that
the M1 strength collected by the scissors mode increases quadratically with the deformation parameter \cite{Ziegler,Heyde}. 

Ref. \cite{Ward12} reported a suppression of the 
 B(M1)$_{out}$ between rotational bands built on different members of the quasineutron j$_{15/2}$ multiplet in $^{235}$U by a factor of 20-50 compared to estimates in the framework of
 the Quasiparticle - Rotor model. In addition, the authors tabulated examples of  B(M1)$_{out}$ values between bands of high-j  multiplet members, which all appear strongly suppressed.
 This systematic quenching of M1 strength suggests that the scissors mode draws M1 strength from the low-energy transitions in analogy to the quenching of the low-energy E1 transitions
  by coupling to the GDR (screening).

Support by the US Department of Energy Grant  No. DE-FG02-95ER40934  is acknowledged. 
Unfortunately Fritz D\"onau passed away before completion of this work. Our community lost a great 
scientist, an enthusiastic researcher, and a friend whom many of us will miss.

\end{document}